# Adsorption of Water on Fluorinated Graphene

Yong Yang[2, 1*], Fuchi Liu[1] and Yoshiyuki Kawazoe[3,4]

1. College of Physics and Technology, Guangxi Normal University, Guilin 541004, China.
2. Key Laboratory of Materials Physics, Institute of Solid State Physics, Chinese Academy of Sciences, Hefei 230031, China.
3. New Industry Creation Hatchery Center (NICHe), Tohoku University, 6-6-4 Aoba, Aramaki, Aoba-ku, Sendai, Miyagi 980-8579, Japan.
4. Department of Physics and Nanotechnology, SRM Institute of Science and Technology, Kattankulathurm, 603203, TN, India.

**ABSTRACT**

In this paper, we investigate the adsorption of water monomer on fluorinated graphene using state-of-the-art first principles methods within the framework of density functional theory (DFT). Four different methods are employed to describe the interactions between water and the carbon surface: The traditional DFT calculations within the generalized gradient approximation (GGA), and three types of calculations using respectively the semi-empirical DFT-D2method, the original van der Waals density functional (vdW-DF) method, and one of its variants. Compared with the adsorption on pristine graphene, the adsorption energies of water on fluorinated graphene are significantly increased, and the orientations of water diploe moment are notably changed. The most stable configuration is found to stay right above the top site of the C atom which is bonded with F, and the dipole moment of water molecule aligns spontaneously along the surface normal.

*E-mail: wateratnanoscale@hotmail.com; yyang@theory.issp.ac.cn

## 1. Introduction

The interaction between water and carbon-based materials such as carbon nanotube, graphite and graphene plays a critical role in the daily life phenomena ranging from wetting, lubrication, heterogeneous ice nucleation and growth, surface catalysis, to the function of carbon nanostructures in biological environment, *etc*. Investigations on this topic have attracted a lot of interests of research in the past decades. For instance, water can exhibit novel phases and anomalous properties when it is confined in carbon nanotube [1, 2]. In the simulation of water in confinement [1] or the super-lubricity of water transport in carbon nanotube [3], the water-carbon interactions play a decisive role, which are usually described by empirical force filed. The parameters of empirical force filed are obtained either by fitting the interatomic potentials from quantum mechanical calculations on small molecules or by reproducing the bulk properties measured by experiments. It remains open about the validity of such empirical force filed in describing the properties of interfacial water at atomic scale. Traditional first-principles calculations based on density functional theory (DFT) are shown to be insufficient to describe the weak water-carbon interactions where the van der Waals (vdW) interactions count [4, 5]. Indeed, the adsorption of water molecules on carbon-based systems serves as a good playground for testing the theory of vdW interactions [4-6].

Upon chemical modifications, graphene and the other recently discovered two-dimensional (2D) materials (e.g., phosphorene, borophene) can exhibit unique properties with comparison to the pristine ones [7-11]. Specially, the adsorption of water can have nontrivial effects on the structural and electronic properties of the underlying 2D material, which in return affects the adsorption structure of water molecules [12-14]. In this work, we study the adsorption of water on fluorinated graphene (abbreviated as F-Gr hereafter), i.e., graphene whose surface decorated with F atoms, to explore the effects of chemical modification on the strength of water-carbon interactions. Fluorination is chosen here by considering the fact that F is the most electronegative element and is expected to induce larger charge perturbations on the graphene substrate with comparison to hydrogenation and oxidation. It is found that the adsorption of water on the fluorinated graphene is enhanced with comparison to the case of water adsorption on pristine graphene [4, 5]. Due to the dipole-diploe interactions between water molecule and the F-Gr substrate, the adsorption geometries are also significantly modified.

## 2. Computational and Modeling Methods

The first-principles calculations are performed using the VASP code [15, 16], which is based on density functional theory (DFT). A plane wave basis set and the projector-augmented-wave (PAW) potentials [17, 18] are employed to describe the electron wave function and the electron-ion interactions, respectively. The exchange-correlation interactions of electrons are described by the PBE type functional [19]. The energy cutoff for plane waves is 600 eV. For structural relaxation and total energy calculations of the $H_2O$/F-Gr system, an $8 \times 8 \times 1$ Monkhorst-Pack k-mesh [20] is generated for sampling the Brillouin zone (BZ). In the description of the van der Waals interactions between water and F-Gr, three types of methods, i.e., the semi-empirical DFT-D2 method of Grimme [21], the original van der Waals density functional (vdW-DF) by Dion *et al.* [22], and the optPBE-vdW (variants of vdW-DF) density functional by Klimeš *et al.* [23, 24], are employed to make a comparison.

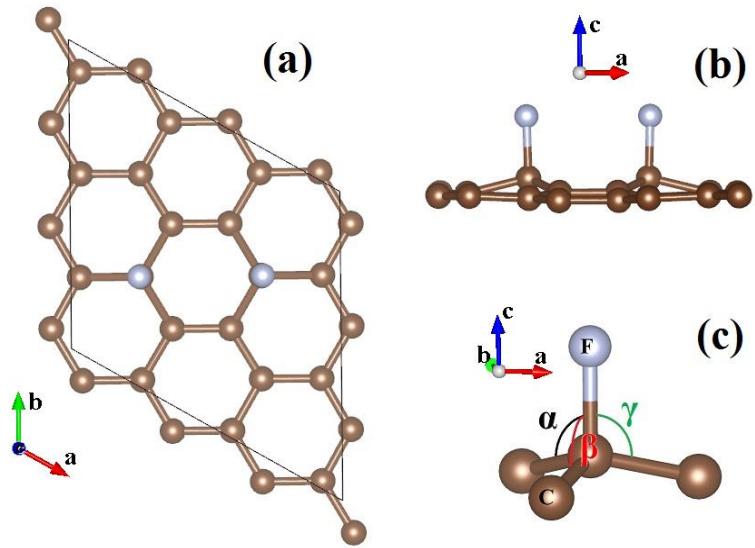

**Fig. 1. (a)** Top view of the unit cell of the fluorinated graphene (F-Gr) sheet. **(b)** Side view of the F-Gr sheet. **(c)** Local C-F bonding geometries of the F-Gr sheet.

The F-Gr substrate on which the water monomers are adsorbed is modeled by decorating two F atoms on the (3×3) supercell of graphene sheet, separated by a vacuum layer of ~ 15 Å in the surface normal (z-direction of supercell) to minimize the artificial interactions of the neighboring images due to periodic boundary condition. The two F atoms are decorated on the top sites of the underlying C atoms, being the third nearest neighbors of each other. Such an adsorption configuration is the most stable configuration for the one-side adsorption of $F_2$ molecule on graphene [25]. The F-Gr substrate is schematically shown in Fig. 1, in which the C atoms which are bonded with F atoms are slightly lifted upward along the z-direction. The geometric parameters describing the F-C local bonding configuration are tabulated in Table 1. Since we are interested in how the water-carbon interactions are affected by the electrical polarity introduced via F decoration, the water monomers are adsorbed on the other side of graphene, to avoid the direct contact and possible bonding of water with F atoms. The adsorption energy of water monomer is calculated as follows:

$$E_{ads} = E[(\text{F-Gr})] + E[(\text{H}_2\text{O})_{isolated}] - E[\text{H}_2\text{O}/(\text{F-Gr})] \qquad (1),$$

where $E[\text{H}_2\text{O}/(\text{F-Gr})]$, $E[(\text{F-Gr})]$, $E[(\text{H}_2\text{O})_{isolated}]$ are respectively the total energies of the adsorption system, the F-Gr substrate, and an isolated water molecule. For the total energy calculation of an isolated water molecule, a (10 Å × 10 Å × 10 Å) supercell is used. Similar formula is applied to the calculation of the adsorption energies of water monomers on pristine graphene, by simply replacing the term $E[(\text{F-Gr})]$ with $E[\text{Gr}]$, and $E[\text{H}_2\text{O}/(\text{F-Gr})]$ with $E[\text{H}_2\text{O}/\text{Gr}]$.

### 3. Results and Discussion

We begin with studying the adsorption of water monomer on pristine graphene. It is found by previous calculations using different types of functionals for the vdW interactions [4, 5], that the configurations with one or two OH bonds of the water molecule pointing downward to the graphene sheet are energetically favored, which have similar adsorption energies and are therefore of similar stability. In this context, either of the two configurations is enough to model the water-graphene interaction, and we choose the configuration with one OH bond pointing vertically towards the graphene surface, which is energetically slightly favored than the configuration with two OH pointing downward [5]. As shown in Fig. 2, it is obvious that the angle $\theta$ formed between the surface normal ($\vec{n}$) and the dipole moment ($\vec{p}$) of the water molecule, is an *obtuse angle* (Table 2). After structural relaxation, the adsorption energies and the parameters describing the adsorption geometries, including the distances of the water molecule to the graphene sheet $Z_O$, and the normal-dipole ($\vec{n}$-$\vec{p}$) angles $\theta$, are listed in Table 2, for the calculations using the four different methods. The adsorption energies and geometries compare well with the data reported in previous works [4, 5]: The adsorption energies is ~ 0.03 eV by PBE, is ~ 0.135 eV by DFT-D2 and ~ 0.147 eV by vdW-DF and ~ 0.152 eV by optPBE-vdW; the height of O atom to the graphene sheet, $Z_O$, varies from ~ 3.3 to 3.5 Å, and the angle $\theta$ varies from ~ 101.5° (vdW-DF) to ~ 122.2° (PBE).

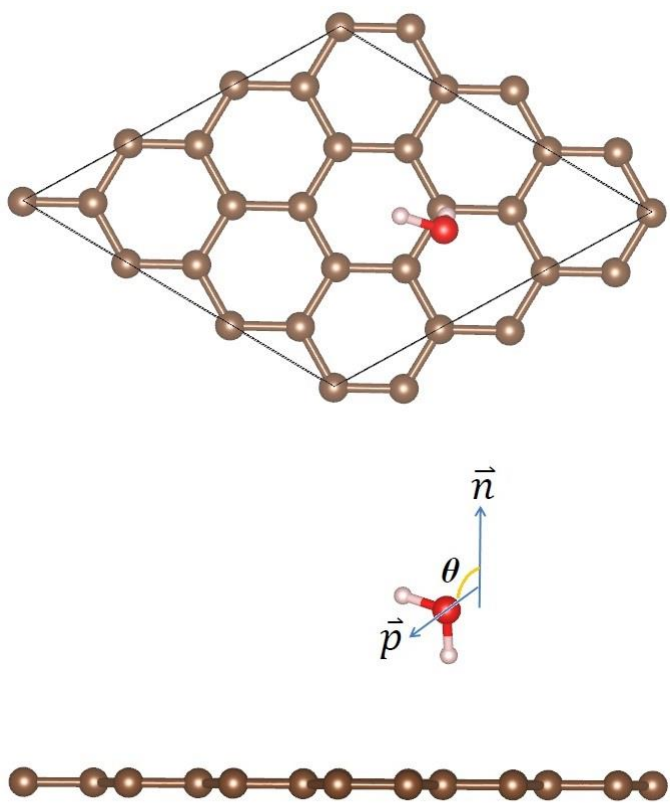

**Fig. 2.** Top view (upper panel) and side view (lower panel) of water monomer adsorption on pristine graphene.

To study the adsorption of water on F-Gr substrate, two situations are considered: The first is that the water molecule stays away from the top sites of the two C atoms bonded with F atoms, and the second is that the water molecule is adsorbed atop one of the C atoms bonded with F atoms on the opposite side of graphene. For each situation, two initial configurations for structural relaxation are examined: the configuration with one OH bond pointing vertically towards the F-Gr substrate (left-upper panels of Figs. 3-4), and the configuration in which two OH bonds pointing away from the F-Gr substrate (left-lower panels of Figs. 3-4).

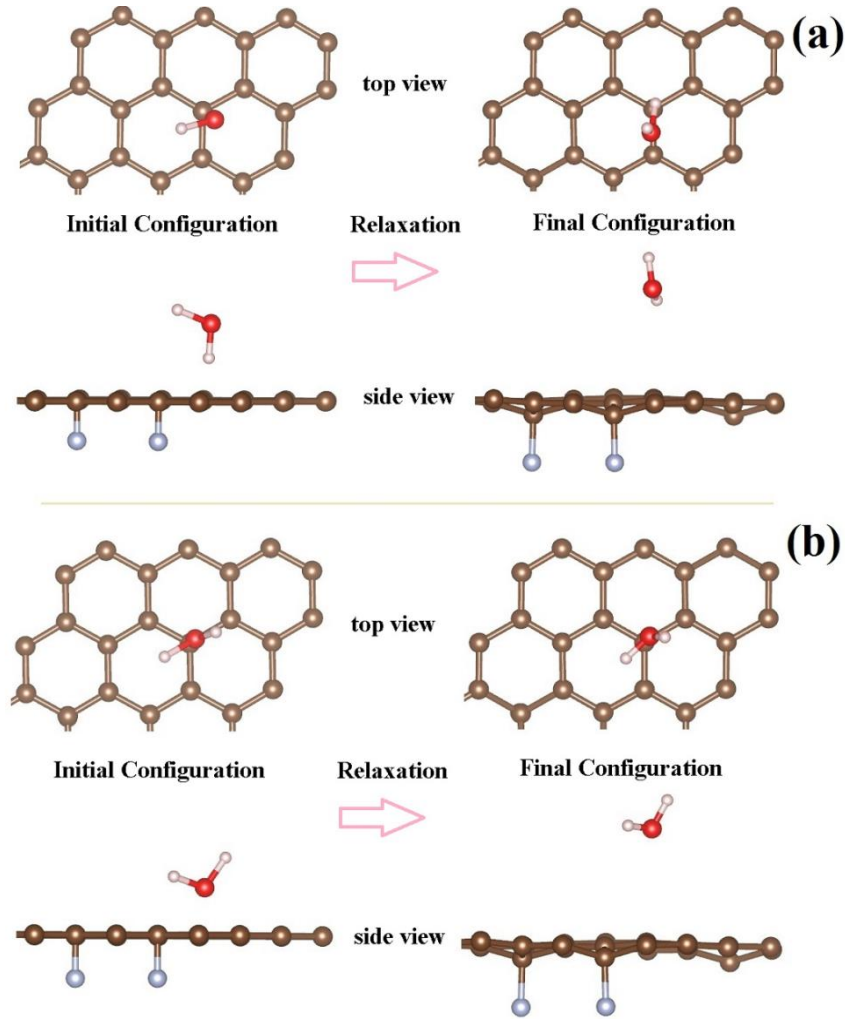

**Fig. 3.** The initial and final configuration of water monomer adsorption on F-Gr sheet, for the situation in which water molecule stays away from the top sites of the C atoms bonded with F. **(a):** Configuration 1 and **(b):** Configuration 2. Note that the C atom which stretches downward (right panels, side view) is actually bonded with one F (not shown to avoid confusion) due to periodic boundary conditions.

In the first situation where there is *no direct contact* between the adsorbed water monomers and the C atoms which are bonded with F (Fig. 3), the initial adsorption configurations are modified after structural relaxation: The configuration shown in Fig. 3(a) (labeled as Configuration 1) is overturned with both OH bonds pointing upwards; while the HOH plane of the configuration shown in Fig. 3(b) (labeled as Configuration 2) are only slightly tilted with comparison to the starting configuration. For both configurations, the O atoms are directed towards the F-Gr substrate after relaxation, with the dipole moments of water forming *acute angles* (Table 3) with the surface normal. Compared with the case of adsorption on pristine graphene, adsorption energy calculated by PBE, vdW-DF and optPBE-vdW methods is notably increased, whereas it remains nearly unchanged for calculations using the DFT-D2 method.

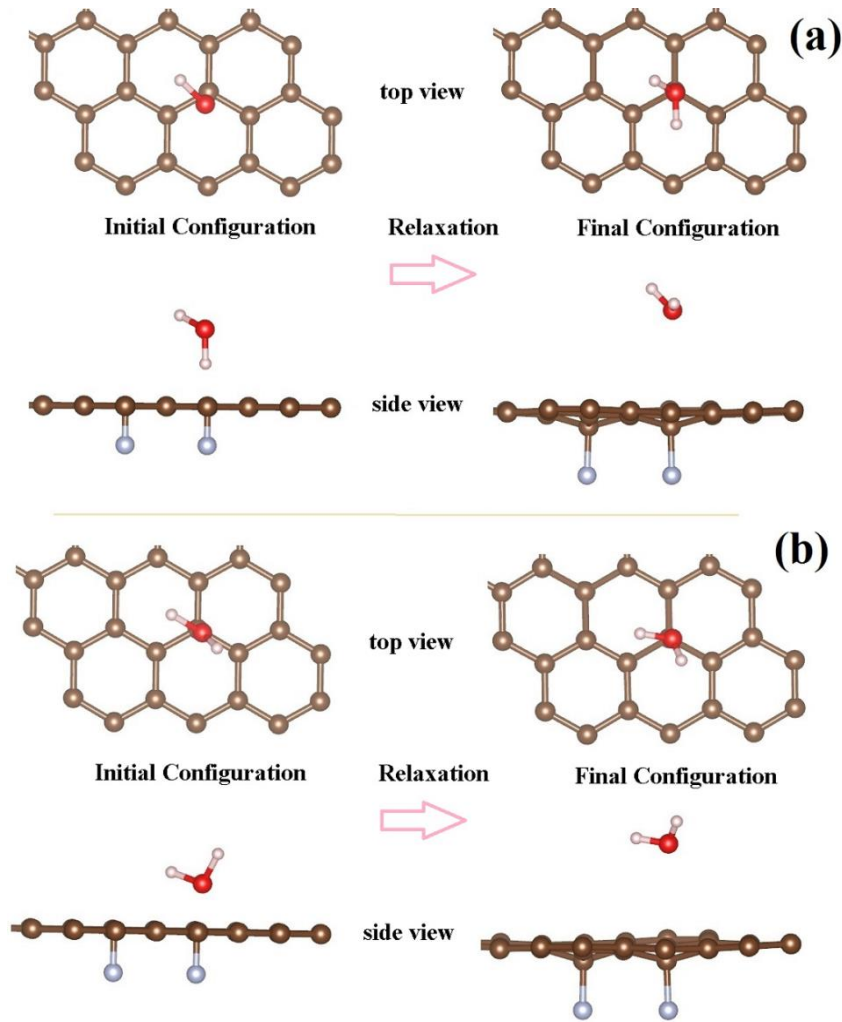

**Fig. 4.** The initial and final configuration of water monomer adsorption on F-Gr sheet,

for the situation in which water molecule stays at the top site of one of the C atoms bonded with F. **(a):** Configuration 3 and **(b):** Configuration 4.

In the second situation where the water monomers are absorbed at the top site of one of the C atoms which are bonded with F (Fig. 4), similar changes in both adsorption configurations are found after structural relaxation: The initial configuration with one OH bond pointing vertically downwards (Fig. 4(a), labeled as Configuration 3) is largely rotated with both OH bonds pointing upwards and the O atom sitting right above the C atom bonded with F; the HOH orientation of the configuration whose two OH bonds initially pointing upwards (Fig. 4(b), labeled as Configuration 4) is only slightly modified. Meanwhile, the adsorption energies calculated by PBE, DFT-D2, vdW-DF and optPBE-vdW methods are all significantly increased for Configuration 3 and 4, as displayed in Fig. 5. The extent of increasing by PBE, DFT-D2, vdW-DF and optPBE-vdW calculation is ~ 96.7%, 14.8%, 36.7%, and 38.8% for Configuration 3, and is ~ 103.3%, 28.9%, 38.1%, and 39.5% for Configuration 4, with reference to the adsorption on pristine graphene, respectively.

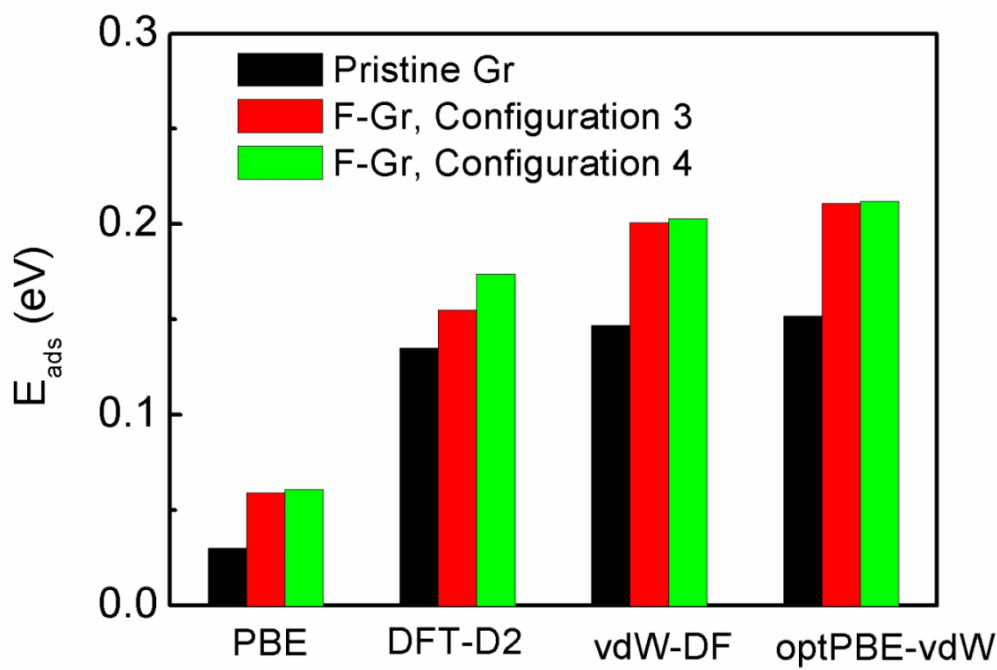

**Fig. 5.** The adsorption energies ($E_{ads}$) of water monomer on pristine and fluorinated graphene (F-Gr), calculated using different types of functionals.

On the other hand, the adsorption energy predicted by DFT-GGA calculation (PBE type functional) is much smaller than the values given by the DFT-D2, vdW-DF and optPBE-vdW methods. This is understandable. Traditional DFT-GGA calculation largely neglects the vdW type interactions due to the semi-local nature of the GGA correlation functional [22]. On the contrary, the other three methods include explicitly the vdW type interactions into the total energies of the system [21-24]. The difference of adsorption energies calculated by the three methods roots in the way of description: DFT-D2 is a semi-empirical method while the vdW-DF and optPBE-vdW incorporate the vdW interactions by modifying the exchange-correlation functional.

As shown in Figs. 1, 3, and 4, the C-F bonding region is obviously distorted in z-direction, and it is difficult to define the planar position of the F-Gr sheet unambiguously. Therefore, we use the distance of O to the nearest C ($d_{OC}$) instead of $Z_O$ to describe the position of water molecule in z-direction. After structural relaxation, the values of $d_{OC}$ for Configurations 1 to 4 range from ~ 3.44 Å to 3.54 Å by PBE, from ~ 3.17 Å to 3.28 Å by DFT-D2, from ~ 3.36 Å to 3.43 Å by vdW-DF, and from ~ 3.12 Å to 3.24 Å by optPBE-vdW.

The energy and geometric parameters obtained from different methods for describing the four adsorption configurations are summarized in Table 3. It is clear that the configurations where the water monomers are adsorbed atop of the C atoms bonded with F atoms, i.e., configurations in the second situation are energetically more stable. For the all the four configurations under consideration, despite the large differences between their initial orientations of the HOH planes, the final dipole moments after relaxation spontaneously form *acute angles* (Table 3) with the surface normal of the F-Gr sheet. Compared with the adsorption configuration shown in Fig. 2, the maximum and minimum of the angle of rotation, $\Delta\theta_{max}$ and $\Delta\theta_{min}$, is respectively ~ 84° and 63° by PBE, ~ 72° and 41° by DFT-D2, ~ 73° and 49° by vdW-DF, ~ 83° and 60° by optPBE-vdW. Such spontaneous alignment of water dipole moments is due to the positively charged graphene sheet upon the decoration of F atoms, where permanent electric dipole moments along the surface normal are induced. The total electric dipole moment along the z-direction (i.e., surface normal) of the F-Gr sheet is calculated to be ~ 0.48 $e \cdot$Å.

Using the Bader charge analysis [26, 27], we are able to calculate the number of electrons intuitively assigned to each atom. For the most stable adsorption configuration, i.e., Configuration 4 (Fig. 4(b)), the number of valence electrons assigned to the two C atoms (labeled as C_1, C_2) and the two F atoms (labeled as F_1, F_2) bonded with them are illustrated in Fig. 6, for different methods. With reference to the charge-neutral state, in which the number of valence electrons is 4 for C and 7 for F (Fig. 6), the net charge assigned to C_1, C_2, F_1, F_2 is respectively ~ +0.52$e$, +0.48$e$, -0.62$e$, -0.62$e$ by PBE, ~ +0.54$e$, +0.55$e$, -0.63$e$, -0.63$e$ by DFT-D2, +0.41$e$, +0.39$e$, -0.57$e$, -0.57$e$ by vdW-DF, ~ +0.42$e$, +0.42$e$, -0.58$e$, -0.58$e$ by optPBE-vdW. The slight difference between the net charges of the two C atoms (C_1, C_2) is due to the weak water-carbon interactions after the adsorption of water monomer atop of one of them.

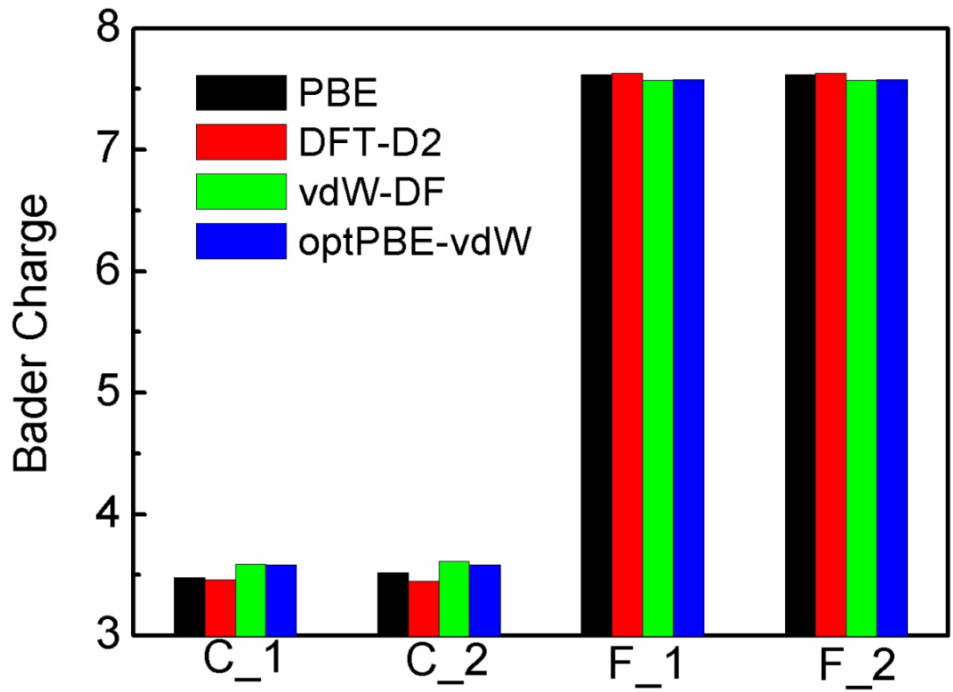

**Fig. 6.** The Bader charge (units: $e^-$) of C_1, C_2 and F_1, F_2 of Configuration 4, calculated using different types of functionals.

To demonstrate this point and to get more insights into the water-carbon interactions, we have further studied the charge density difference of Configuration 4, before and after the adsorption of water monomer. The charge density difference is calculated using the following formula:

$$\Delta\rho = \rho[H_2O/(F\text{-}Gr)] - \rho[(F\text{-}Gr)] - \rho[(H_2O)_{isolated}] \qquad (2),$$

where $\rho[H_2O/(F\text{-}Gr)]$, $\rho[(F\text{-}Gr)]$, and $\rho[(H_2O)_{isolated}]$ are respectively the charge densities of the adsorption system, the F-Gr substrate, and an isolated water molecule. The results are schematically shown in Fig. 7. From the spatial distribution of $\Delta\rho$, one sees clear deviation from the three-fold rotation symmetry of underlying F-Gr lattice. This is the consequence of the enhanced water-carbon interactions upon F decoration.

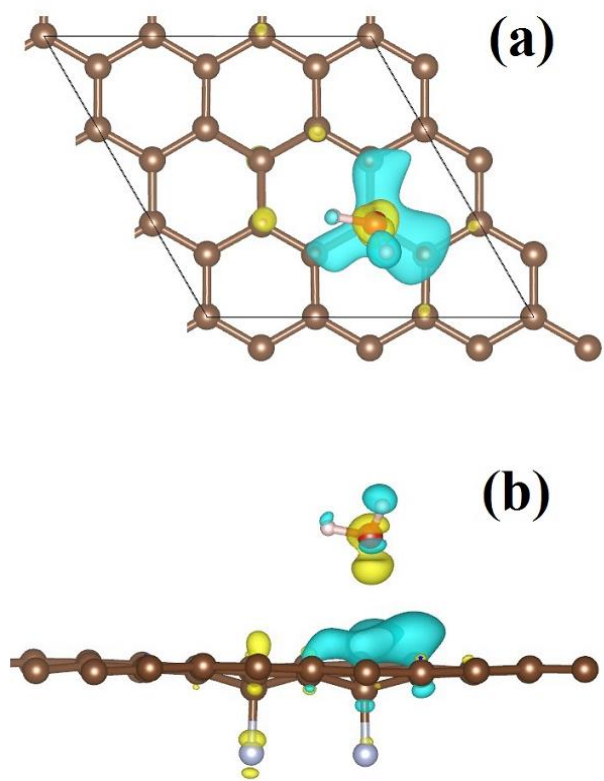

**Fig. 7.** Top view (upper panel) and side view (lower panel) of the charge density difference of the water monomer adsorption on F-Gr (Configuration 4). The isovalue of plotting the surfaces is 0.0004 $e$/(Bohr$^3$).

Then it is natural to ask which type of the vdW methods better describes the water-(F-Gr) interactions. From the calculation of adsorption energy of water monomer on pristine graphene, the DFT-D2 method performs slightly better than the other methods when comparing with the data given by random phase approximation (RPA) and diffusion Monte Carlo (DMC), which is ~ 0.07 to 0.08 eV [4, 5]. However, the differences (< 0.02 eV) of adsorption energies given by the three methods (DFT-D2, vdW-DF, optPBE-vdW) are within the error bar of calculations. The DFT-D2 method treats the vdW interactions by adding a semi-empirical atomic-pairwise potential term (in the form of $C_6/R^6$) to the conventional Kohn-Sham DFT total energy [21]. Both the vdW-DF and its variant optPBE-vdW deal with the vdW interactions by including the long-range part of the correlation interaction (nonlocal correlation) into the exchange-correlation functional [22-24]. Despite the difference in the GGA-type exchange functional [22-24], the vdW-DF and optPBE-vdW predict essentially the same adsorption energies and geometries for water adsorption on F-Gr (Table 3). In the sense of "*ab initio*" calculations, vdW-DF and optPBE belong to higher-level approaches which employ fewer empirical parameters and are logically more self-consistent with comparison to DFT-D2. To find out which one is better, one needs the data of experimental measurement; or the data obtained from the more accurate theoretical methods such as RPA or DMC, whose huge computational burden is, however, beyond our affordable supercomputing resource. Nevertheless, experimental data should be the ultimate benchmark to figure out the best theoretical method. Despite the numerical differences, the following trend is the same: 1) Enhancement of adsorption energies (Fig. 5); 2) Reorientation of water dipole moment (Figs. 3-4) with comparison to the adsorption on pristine graphene. Both can be understood from the simple picture of electrostatic interactions: The induced dipole moments upon F decoration lead to the enhanced water-substrate interactions and reorientation of water diploe moment. Similar results can be expected when F atoms are replaced by O atoms except that the magnitude of enhancement in adsorption energy would be smaller due to the lower electronegativity of O. In the case of hydrogenation, much weaker effects on water dipole moment are expected due to the much lower electronegativity of H with comparison to F and O. Compared with adsorption on pristine graphene, the orientation of water dipole moment on hydrogenated graphene would not change due to the fact that H is less electronegative than C.

To further demonstrate the effects of electric dipole moment on water adsorption, we have investigated the evolution of total energies of the adsorption systems during the process of structural optimization. The results of DFT-PBE calculations are shown in Fig. 8. Significant fluctuations and differences are found for the initial relaxation steps (Nstep ~ 10). For instance, the total energy of the starting point (Nstep = 1) of Configuration 1 and Configuration 2 differs by ~ 0.6 eV, due to the large difference of the starting configurations (e.g., orientations of HOH dipole plane, $H_2O$-Gr distances). In spite of such differences, the dipole-dipole interactions between water molecule and the underlying fluorinated graphene surface lead to good convergence between the total energies of the two initially different adsorption configurations after ~ 40 steps of structural relaxation (Fig. 8). Similar convergence behavior is observed for the calculations using DFT-D2, vdW-DF, and optPBE-vdW. This trend is also evidenced by the adsorption energies and geometric parameters tabulated in Table 3. The adsorption energies of the same adsorption situation converge to within 5 meV for relaxations using the three fully self-consistent methods (PBE, vdW-DF and optPBE-vdW), while slight divergence is observed for results given by DFT-D2 due to the semi-empirical nature of this method.

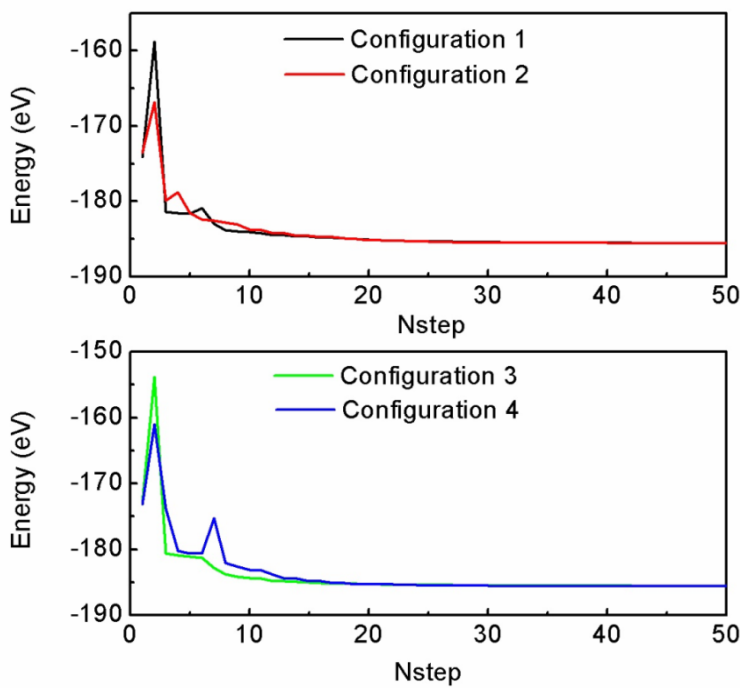

**Fig. 8**. Calculated total energies of the water-(F-Gr) systems, as a function of the number of steps of structural relaxation (Nstep).

## 4. Summary

In conclusion, we have examined the adsorption of water monomer on fluorinated graphene using first-principles methods, including traditional DFT-GGA calculations, the semi-empirical DFT-D2 method and the van der Waals density functional and one of its variants to treat the van der Waals interactions between water and carbon surface. The water-carbon interactions are found to be significantly enhanced and the dipole moments of the adsorbed water molecules align spontaneously with the surface normal direction of the F-Gr sheet, along with the existing strong electric dipole moments due to F decoration. The significant enhancement of adsorption energies indicates that, the wetting properties of water on graphene surface can be manipulated by fluorination. For instance, by fluorination and de-fluorination, the graphene surface may be switched between hydrophobic and hydrophilic, which can be tested by measuring the contact angles of water droplets. Our work demonstrates the possibility of tuning the strength of water-graphene bonding with F-decoration, and thus the strength of friction and lubrication at nanoscale.

**Acknowledgments**

This work is jointly supported by the National Natural Science Foundation of China (No. 11664003, 11474285), Natural Science Foundation of Guangxi Province (No. 2015GXNSFAA139015), and the Scientific Research and Technology Development Program of Guilin (No.2016012002). We gratefully acknowledge the crew of Center for Computational Materials Science of the Institute for Materials Research, Tohoku University for their continuous support of the SR16000 supercomputing facilities. We also thank the staff of the Hefei Branch of Supercomputing Center of Chinese Academy of Sciences for their support of supercomputing resources.

**Table 1.** The geometric parameters describing the fluorinated graphene for water adsorption: C-F bond lengths ($L_{CF}$), the equilibrium distance of the decorated F atoms (labeled as F_1 and F_2) with reference to the pristine graphene surface ($Z_F$), and the angles (α, β, γ) formed between the adsorbed F and the C atoms nearby (Fig. 1).

| Atom | $L_{CF}$ (Å) | $Z_F$ (Å) | α (°) | β (°) | γ (°) |
|---|---|---|---|---|---|
| F_1 | 1.48 | 1.90 | 104.33 | 104.53 | 102.50 |
| F_2 | 1.48 | 1.90 | 104.52 | 104.49 | 102.39 |

**Table 2.** The adsorption energy ($E_{ads}$), the equilibrium distance of the O atom in water molecule to pristine graphene surface ($Z_O$), and the angle ($\theta$) between the electric dipole moment of water molecule and the surface normal of graphene sheet (named as diploe-surface normal angle hereafter), as illustrated in Fig. 2.

| Calculation Approach | Configuration in Fig. 2 | | | Reference data1 [4] | | Reference data2 [5] | |
|---|---|---|---|---|---|---|---|
| | $E_{ads}$ (eV) | $Z_O$ (Å) | $\theta$ (°) | $E_{ads}$ (eV) | $Z_O$ (Å) | $E_{ads}$ (eV) | $Z_O$ (Å) |
| PBE | 0.030 | 3.57 | 122.21 | 0.031 | 3.65 | 0.032 | 3.607 |
| DFT-D2 | 0.135 | 3.33 | 114.07 | --- | --- | --- | --- |
| vdW-DF | 0.147 | 3.57 | 101.53 | --- | --- | 0.133 | 3.533 |
| optPBE-vdW | 0.152 | 3.32 | 108.92 | --- | --- | 0.159 | 3.400 |

**Table 3.** Calculated adsorption energy ($E_{ads}$), the equilibrium distance of O atom to the nearest C atom of the fluorinated graphene ($d_{OC}$), and the dipole-surface normal angle ($\theta$) of the four adsorption configurations of water monomers.

| Calculation Approach | Configuration 1 | | | Configuration 2 | | | Configuration 3 | | | Configuration 4 | | |
|---|---|---|---|---|---|---|---|---|---|---|---|---|
| | $E_{ads}$ (eV) | $d_{OC}$ (Å) | $\theta$ (°) | $E_{ads}$ (eV) | $d_{OC}$ (Å) | $\theta$ (°) | $E_{ads}$ (eV) | $d_{OC}$ (Å) | $\theta$ (°) | $E_{ads}$ (eV) | $d_{OC}$ (Å) | $\theta$ (°) |
| PBE | 0.047 | 3.52 | 59.57 | 0.048 | 3.44 | 38.16 | 0.059 | 3.54 | 44.60 | 0.061 | 3.49 | 44.74 |
| DFT-D2 | 0.130 | 3.22 | 61.66 | 0.131 | 3.28 | 70.22 | 0.155 | 3.18 | 42.18 | 0.174 | 3.17 | 73.23 |
| vdW-DF | 0.172 | 3.39 | 52.92 | 0.169 | 3.36 | 32.12 | 0.201 | 3.43 | 28.68 | 0.203 | 3.40 | 33.54 |
| optPBE-vdW | 0.167 | 3.21 | 48.89 | 0.164 | 3.12 | 31.61 | 0.211 | 3.18 | 28.55 | 0.212 | 3.24 | 26.06 |